\documentclass[fleqn,usenatbib]{mnras}

\usepackage[T1]{fontenc}

\DeclareRobustCommand{\VAN}[3]{#2}
\let\VANthebibliography\thebibliography
\def\thebibliography{\DeclareRobustCommand{\VAN}[3]{##3}\VANthebibliography}

\usepackage{graphicx}	
\usepackage{amsmath}	
\usepackage{amssymb}	

\usepackage{color}
\usepackage[dvipsnames]{xcolor}
\usepackage{ulem}
\DeclareGraphicsExtensions{.png,.pdf,.eps,.jpg,.tiff,.ps}

\newcommand{\vunit}{\mbox{\,km\,s$^{-1}$}}

\newcommand{\Msun}{\mbox{\,$M_\odot$}}
\newcommand{\Lsun}{\mbox{\,$L_\odot$}}
\newcommand{\Rsun}{\mbox{\,$R_\odot$}}

\newcommand{\mic}{\mbox{$\,\mu$m}}

\usepackage{graphicx}	
\usepackage{amsmath}	
\usepackage{amssymb}	
\usepackage{pdflscape}
\usepackage{ulem}
\definecolor{cyan(process)}{rgb}{0.0, 0.72, 0.92}

\newcommand{\fion}[2]{\mbox{[\ion{#1}{#2}]}}

\newcommand{\cmfgen}{\hbox{\sc cmfgen}}
\newcommand{\mf}[2]{{\ensuremath{Y_{\hbox{\scriptsize #1}} = #2}}}
\newcommand{\Rstar}{\hbox{$R_*$}}
\def\blankhalf{\par\vskip 4pt\noindent}
\newcommand{\vinf}{\hbox{$V_\infty$}}

\newcommand{\Mdot}{\hbox{$\dot{M}$}}
\newcommand{\Msunyr}{\hbox{$M_\odot$\,yr$^{-1}$}}
\newcommand{\kms}{\hbox{km\,s$^{-1}$}}

\title[Infrared spectrum of Pa30]
{Infrared spectrum of the Type Iax supernova remnant Pa30}
\author[Banerjee et al.]
{D. P. K. Banerjee$^{1}$\thanks{E-mail: dpkb12345@gmail.com},
A. Evans,$^{2}$\thanks{E-mail: a.evans@keele.ac.uk. Corresponding author.},
S. Starrfield$^3$,
C. E. Woodward$^4$,
T. R. Geballe$^5$, \newauthor
V.  Joshi$^1$, 
{D. J. Hillier$^6$,}
{L. Izzo$^7$,}
{M. Della Valle$^{8}$,}
{G. Shaw$^{9}$,} 
{G. Rangwal$^{10}$,}
{S. Sharma$^{10}$}
 \\ \\
$^{1}$Physical Research Laboratory, Navrangpura,
Ahmedabad, Gujarat 380009, India \\
$^{2}$Astrophysics Research Centre, Lennard Jones Laboratory,
Keele University, Keele, Staffordshire,  ST5 5BG, UK\\
$^{3}$School of Earth and Space Exploration, Arizona
State University, Box 876004, Tempe, AZ 85287-6004, USA \\
$^4$Minnesota Institute for Astrophysics, University of Minnesota, 116 Church Street SE, Minneapolis, MN 55455, USA \\
$^5$Gemini Observatory/NSF's NOIRLab, 670 N. A`ohoku Place, Hilo, HI 96720 USA\\
$^6$Department of Physics and Astronomy \& Pittsburgh Particle
Physics, Astrophysics, and Cosmology Center (PITT PACC), \\
University of Pittsburgh, Pittsburgh, PA 15260, USA \\
$^7$INAF, Osservatorio Astronomico di Capodimonte, Salita 
Moiariello 16, I-80131 Napoli, Italy\\
$^8$INAF–Osservatorio Astronomico di Padova, Vicolo dell’Osservatorio 5, 35122 Padova, Italy \\
$^{9}$Department of Astronomy and Astrophysics, Tata Institute of
Fundamental Research, Mumbai 400005, India \\
$^{10}$Aryabhatta Research Institute of Observational Science,  Manora Peak, Nainital 263002, India\\
} 

\date{Accepted XXX. Received YYY; in original form ZZZ}

\pubyear{2026}

\begin{document}
\label{firstpage}
\pagerange{\pageref{firstpage}--\pageref{lastpage}}
\maketitle

\begin{abstract}
The central star of Pa30 and its nebulosity offer one of
the best opportunities to study a SN Iax explosion and
its remnant. Pa30 is thought to be the product of the
historical supernova explosion of 1181. No infrared spectroscopy
of its extremely hot  central star has hitherto been reported.
Here, we present a 0.75--5\mic\ spectrum taken by the
Spectro-Photometer for the History of the Universe, 
Epoch of Reionization and Ices Explorer (SPHEREx), together
with ground-based infrared spectra, at higher resolution,
covering the 0.8-2.5\mic\ region. The spectra contain very broad
lines, which we have attempted to identify and model
using the codes CMFGEN and CLOUDY. 
The CMFGEN modelling suggests that most of the lines
are due to \ion{O}{vi}. However, CLOUDY modelling, using
expected theoretical abundances for the bound remnants of SN Iax
explosions, provides evidence for
the presence of coronal lines of Si, S, Mg, and Al arising
from a shocked gas. This strengthens the suggestion that Pa30
is indeed the remnant of a SN Iax explosion.
\end{abstract}




\section{Introduction}

Patchick 30 (hereafter Pa30; also known as IRAS 00500+6713,
2MASS J00531123+6730023, IPHAS J005311.21+673002.1,
WISE J005311.24+673002.4, Gaia DR3 526287189172936320)
has been identified as the remnant of the ``guest star''
widely observed in China, Japan and Egypt in 1181 August
\citep{kronberger2014,ritter21,fischer25}. 
{It was first identified as the remnant of a putative 
SN Iax explosion by \cite{oskinova20}, using X-ray data, and}
data from the Wide-Field Infrared Survey Explorer 
\citep[WISE;][]{wise}. Earlier optical spectroscopy
\citep{gvaramadze19} revealed emission
lines of \ion{O}{vi} and \ion{C}{iv}; these authors also
concluded that the stellar remnant has a luminosity
$10^{4.6}$\Lsun\ and a high effective temperature
(211,000~K). \citeauthor{gvaramadze19} suggested that the 
properties of Pa30 
{and its nebula are consistent with the merger 
of two massive white dwarfs (WDs) that resulted in a
super-Chandrasekhar-mass remnant. However, 
\cite{lykou23} argue for a final mass in the range
1.2--1.45\Msun.} The widths of the 
\ion{O}{vi} lines suggest a wind velocity of 16,000\vunit,
likely magnetically driven  \citep{gvaramadze19}.
{\citeauthor{gvaramadze19} deduced a magnetic field of
$2\times10^8$~G; however, \cite{lykou23} deduced an
upper limit of $2.5\times10^6$~G, based on the absence
of Zeeman splitting in the \ion{O}{viii} lines.}

{\cite{ritter21} reported high velocities in
the nebulosity, consistent with an origin in the 1181
event. \cite*{fesen23} found that Pa30 is surrounded by a
firework-like nebulosity of radius $\simeq90\arcsec$,
with remarkably narrow, radial, filaments that 
show [\ion{S}{ii}] emission. These radial filaments
are revealed to be a sequence of knots or clumps
\citep{cunningham26}. The nebula is reported
as being rich in sulphur and argon,
and deficient in hydrogen and helium \citep{fesen23}.}

\begin{figure*}
\centering
 \includegraphics[width=.7\textwidth,keepaspectratio]{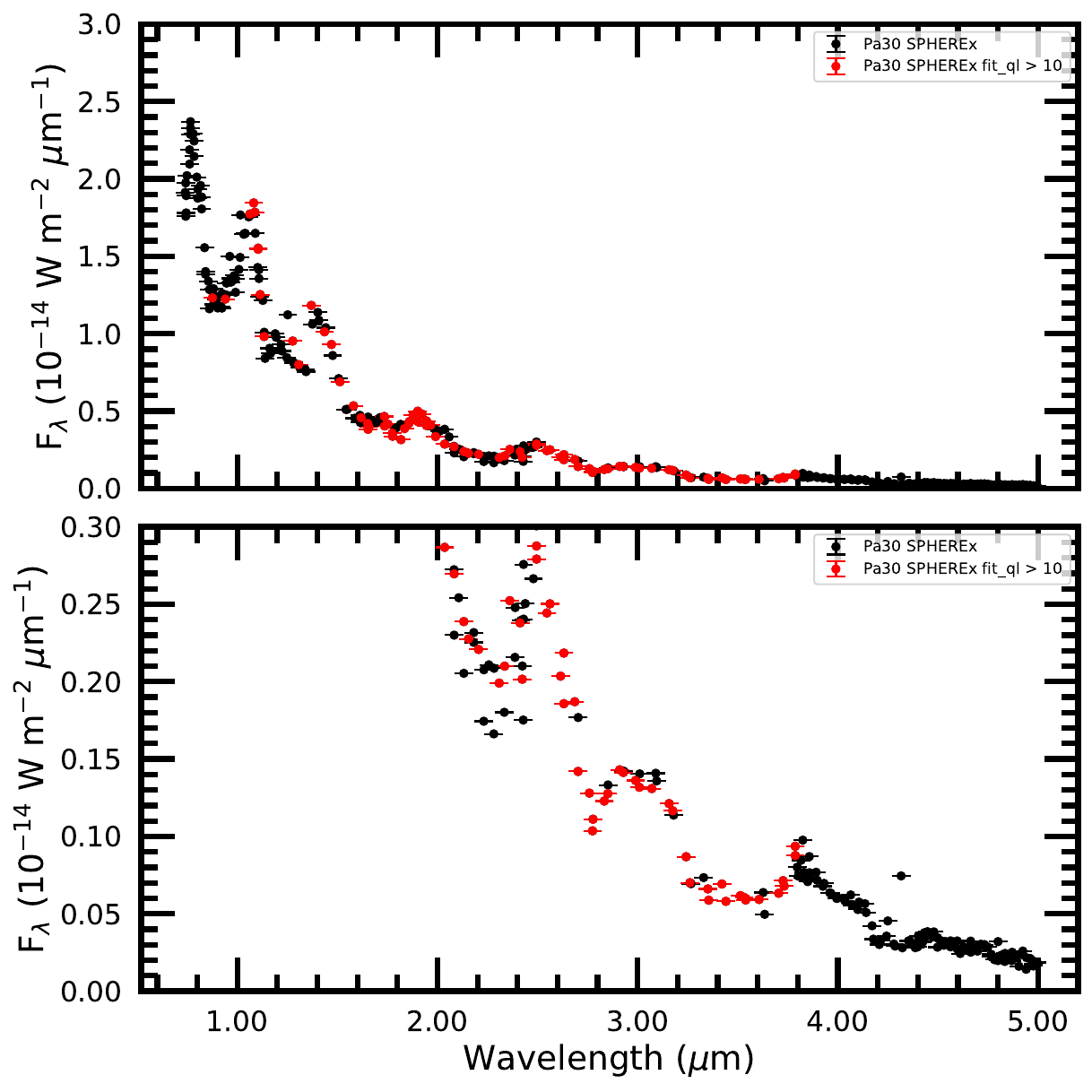}
 \caption{Top: SPHEREx spectrum.
 Bottom: zoomed section of the longer wavelength region of the
 spectrum. In both panels, the red points are those with
 fit\_ql $>10$.  See text for details.\label{pa30}} 
\end{figure*}

\begin{figure*}
\centering
 \includegraphics[width=.65\textwidth,keepaspectratio]
 {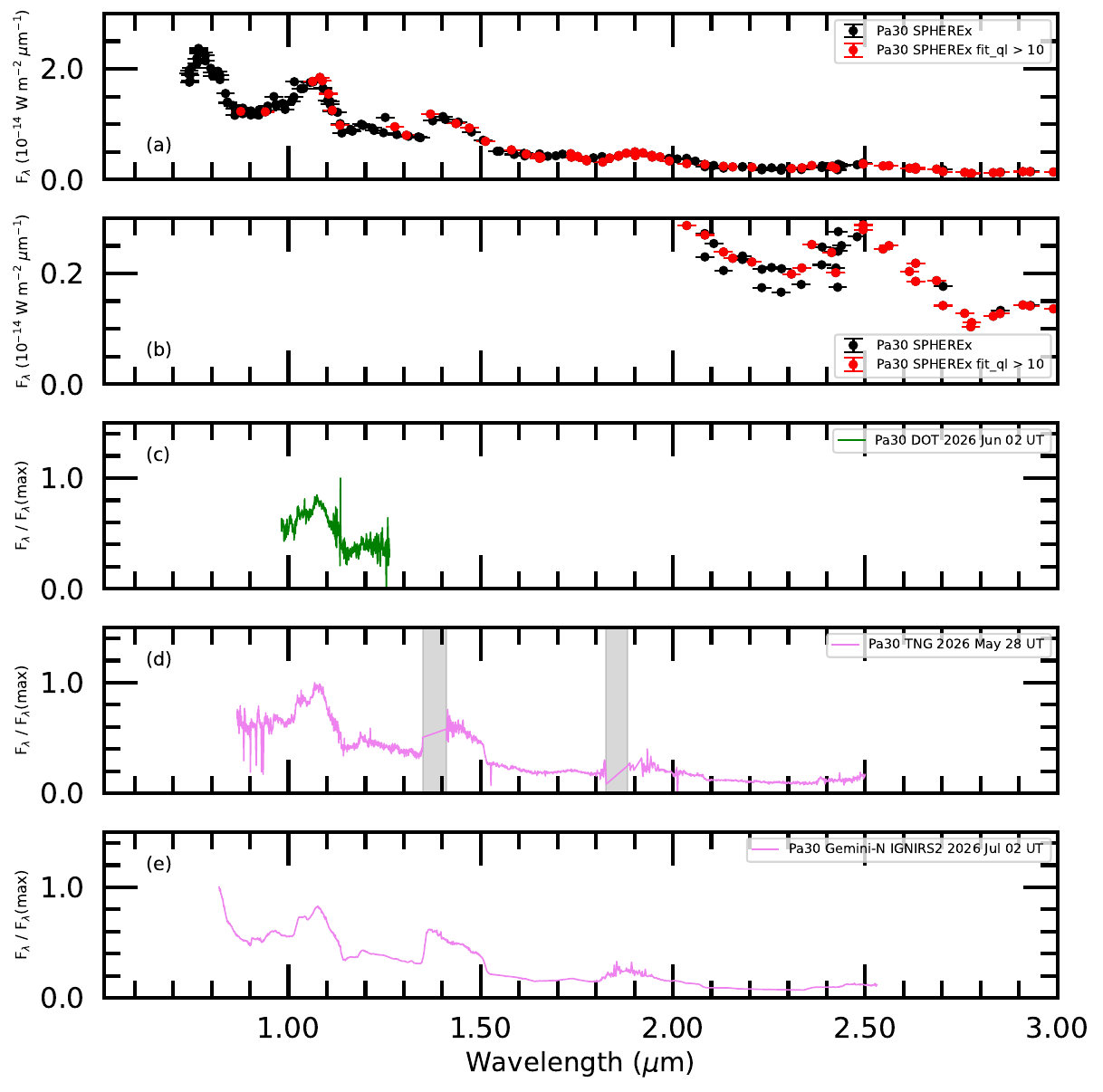}
  \caption{(a) SPHEREx 0.7--3.0\mic\ spectrum; (b) SPHEREx
  2.0--3.0\mic\ expanded vertically by a factor of 10. 
  Colours of points as in Fig.~\ref{pa30}.
  (c) DOT data. Only part of the spectrum is shown as other
  cross-dispersed orders were too noisy.
  (d) TNG data.
  (e) Gemini North spectrum; in most of the spectrum the 
  error bars are too small to be visible.
 The gray blocks indicate regions of poor
 atmospheric transmission.
 \label{ground}} 
\end{figure*}

It has been proposed that the Pa30 SNR originated from a
supernova (SN) of Type Iax \citep{oskinova20}.
This is a subtype of Type Ia SNe that results from a
failed deflagration that does not completely unbind the WD.
Instead of completely
destroying the WD, the explosion stutters, leaving behind a
surviving, extremely hot, central star.
Type Iax SNe are characterized by low peak luminosities
{\citep[absolute visual magnitudes 
$\sim-14.2$ to $-18.9$;][]{foley13}}, and slow
expansion velocities at maximum light 
\citep[$\sim2,000-8,000$\vunit;][]{foley13}. Further,
they do not display the near-infrared secondary maximum
in their light curves that is seen in normal Type SNe Ia
\citep{foley13,jha17}.

{Recent observations of very faint SNe Iax, 
such as SN~2024vjm \citep{zimmerman26},
further support the idea that weak thermonuclear
explosions on WDs can eject only a small amount
of mass, and leave behind a gravitationally bound remnant.
Although SN~2024vjm and Pa30 are observed at very
different evolutionary stages, this result 
is consistent with the Iax interpretation for Pa30.}

We present here an infrared (IR; 0.75--5.0\mic)
spectrum of the central star of Pa30, obtained
with the Spectro-Photometer for the
History of the Universe, Epoch of Reionization, and Ices Explorer
\citep[SPHEREx;][]{crill20,bock26}
that provides an improved understanding of the nature of Pa30.
{These data are supplemented by a ground-based 
0.9--2.5\mic\ spectrum from the 3.6~m Telescopio Nazionale Galileo
(TNG), a 0.7--2.5\mic\ spectrum from the Devasthal
Optical Telescope (DOT),
and a 0.8--2.5\mic\ spectrum from the Gemini North Telescope.}

\section{Observations}

\subsection{SPHEREx}

The spectrum was taken by SPHEREx in 102 colour bands
covering the wavelength range 0.75--5.0\mic. The 
Short-Wavelength Channel (0.75\mic--2.44\mic) covers 41
bands at a resolution of $R\sim41$, while the Long-Wavelength
Channel (2.40\mic--5.01\mic) covers approximately 61 bands
at $R\sim35$; {the pixel scale is 6\farcs2. Details
are given in \cite{bock26}}.
The observations were carried out between MJD 60882.40
and 61073.3 (2025-07-25 to 2026-02-01).
The SPHEREx spectrum is shown in Fig.~\ref{pa30}.
Some of the spectrophotometry data do not meet the 
SPHEREx criterion that the fit quality metric
``fit\_ql'' value
\begin{equation}
 \mbox{fit}\_{\rm ql} = \left \langle \frac{|p_k - m_k|}{\sigma_k} \right \rangle 
\end{equation}
be close to unity. Here $p_k$ is the pixel value,
$m_k$ is the best estimate of the source flux,
and $\sigma_k$ is the 
per-pixel uncertainty. Some of the data have 
fit\_ql values up to $\sim30$, and points having
fit\_ql $\ge10$ are flagged in Fig.~\ref{pa30}.
{We visually inspected each of the 102
SPHEREx narrow band spectrophotometric frames for signs of 
extended line emission \citep[e.g., corresponding to the
elongated filamentary structures seen in
\fion{S}{ii};][]{fesen23}, or emission from dust. No such
emission was detected. Accordingly the  SPHEREx data
discussed here refer only to the central star.}

\subsection{{Gemini North}}

Near-IR (0.8--2.5\mic) spectra of Pa30 were obtained at the
Frederick C. Gillett Gemini North Telescope on Maunakea in
photometric conditions on UT 2026 July 2 for program 
GN-2024A-FT-213. The facility instrument GNIRS
\citep[GNIRS;][]{elias06} was used in its normal
cross-dispersed mode using the 32$\ell$~mm$^{-1}$
grating and the 0\farcs45 arcsec slit aligned along
the mean parallactic
angle. The data were obtained in the standard ABBA
nod-along-slit mode with a nod distance of 3\farcs0.
The total exposure time was 1920~seconds, and the mean
airmass was 1.64. The telluric A0V standard star HD 8424 was
observed in the same manner at an airmass of 1.71 directly after
observations of the nova were obtained. 

Data reduction used a combination of IRAF 
\citep{tody86} and Starlink Figaro \citep{curry14}
as described, for example, in \cite{evans22}.
The coadded spectra were reduced order by order, and then stitched
together after making small scaling adjustments so that
the spectra matched where the orders overlapped. 

The nominal resolving power ($\lambda/\delta\lambda$)
of the observed spectrum is 1,200. The final flux-calibrated
spectrum of the Pa30 presented here is plotted in 0.001\mic\
bins in the 0.82--1.50\mic\ interval, in 0.0015\mic\
bins in the 1.50-1.90\mic\ interval, and 0.002\mic\
bins in the 1.90--2.53\mic\ interval (Fig.~\ref{ground}(e)).

\subsection{{Telescopio Nazionale Galileo (TNG)}}
A 0.9–2.5\mic\ spectrum of Pa30 was obtained
using the Near Infrared
Camera Spectrometer (NICS) of the TNG using the $I\!J$ and
$H\!K$ grisms, which cover 0.9--1.45\mic\ and 1.4--2.50\mic, 
respectively. A $1''$ wide and $4'$ long slit was used, which
gives a resolving power of $R = 500$. Pa30 and a B9V standard
star (HIP\,6820) were observed in an ABBA sequence. Pa30 was
observed for a total integration time of 1200~s in each
of $I\!J$ and $H\!K$, whereas the standard star was observed
for 80~s and 200~s in $I\!J$ and $H\!K$, respectively. The mean
observation time and airmass for Pa30 were 2026 May 28.17 UT and 2.1,
respectively. For the standard star, they were 
2026 May 28.21 UT and 2.04, respectively. An argon lamp was used
for wavelength calibration. The spectra were reduced in the conventional
manner \citep[e.g.,][]{banerjee14} using Python routines
(Fig.~\ref{ground}(d)).

\subsection{{Devasthal Optical Telescope (DOT)}}
A 0.7--2.5\mic\ cross-dispersed spectrum ($R\sim1600$) was
obtained on 2026 May 27.90 UT using the TIFR ARIES Near-IR 
Spectrometer \citep[TANSPEC;][]{sharma22} mounted on the 
DOT. The source and
the A0V standard star (HD\,2904) were observed in two dithered
positions along the slit. The total integration time for the
source was 2715~s (with individual exposures of 147.36~s);
the airmass at the beginning and end of the observations was
1.99 and 1.60, respectively. The standard star was observed
twice, before and after the target source, at a similar airmass.
Argon, neon, and continuum lamps were also taken during the
observations.  Data reduction was done using Python routines
(Fig.~\ref{ground}(c)).

{The ground-based spectra are shown in Fig.~\ref{ground},
together with the SPHEREx data.
As is evident, there is good ageement between the
ground based and SPHEREx data, vindicating 
the reality of the extremely
broad (Full Width at Zero Intensity (FWZI) $\sim3\times10^4$\vunit)
features seen in the latter.}

\section{{Modelling}}

In order to better understand these NIR spectra
of Pa30 we have undertaken modelling using CMFGEN and 
CLOUDY. The data were dereddened using the extinction law of
\cite*{cardelli89}, with  $E(B-V)=1.05$ and $R=3.1$;
a distance of 2.1~kpc was used \citep{lykou23}. 

\subsection{{{\Large CMFGEN modeling of Pa30}}}

To model the spectra of Pa~30, we used the non-LTE radiative
transfer code, \cmfgen\ \citep{hillier98}, which has been 
used to model a wide variety of stars and supernovae.
This code solves for the non-LTE level populations by
solving the radiative transfer in the comoving-frame.
An iterative technique is used to achieve consistency
between the level populations and radiation field.
Most of the modelling presented here uses a 
version of \cmfgen\ which has been modified to use 
Message Passing Interface (MPI)
rather than Open-MPI (Hillier et al., in preparation).

For the initial model we adopted  parameters similar to
that used in the study of \cite{fesen23};
the essential parameters are listed in Table~\ref{params}.
The atmosphere of this star is primarily composed of
carbon (mass fraction \mf{C}{0.13}), oxygen
(\mf{O}{0.42}) and neon (\mf{Ne}{0.44}). Other
species have solar mass fractions except for hydrogen
(\mf{H}{0}) and helium (\mf{He}{0.017}). The high mass
fractions of C, O, and Ne are needed to explain the
strengths of emission features in both the ultraviolet (UV)
and optical, while the He mass fraction is not well constrained.
The relative strengths of \ion{O}{v} and \ion{O}{vi}
emission features, the equivalent widths of \ion{O}{vi}
$\lambda\lambda$ 3811, 3834\AA, and the absolute flux level,
provide the primary modeling constraints. Typical fits to the UV
and optical spectra, together with line identifications,
can be seen in \cite{lykou23}.

\begin{table}
\centering
\caption{CFMGEN model summary.\label{params}}
\blankhalf
\begin{tabular}{llll}   
\hline
\hline
 \vspace{-0.3cm}\\
 Parameter  &   Value \\
 \hline
 $L$         &   $3.6 \times 10^4$ \,\Lsun \\
 \Rstar($\tau=2.3$)  & 0.136\, \Rsun \\
 \Mdot   &  $2.0 \times 10^{-6}$\,\Msunyr \\
 \vinf     &  15,000\,\kms \\
 $Y_{\hbox{\scriptsize He}}$   &   0.017  \\
  $Y_{\hbox{\scriptsize C}}$  &   0.13 \\
  $Y_{\hbox{\scriptsize O}}$  &   0.41   \\
 $Y_{\hbox{\scriptsize Ne}}$  &   0.44  \\
\hline
\end{tabular}
\end{table}

To more accurately model low lying transitions between terms of
the same parity that occur in the IR, we updated our 
model atoms to include several additional ionisation stages
(\ion{Mg}{v}, \ion{Si}{vii}, \ion{Si}{viii}, \ion{S}{ix})
their relevant transition probabilities, and 
their collision strengths.
These data were take from the NIST atomic data
base\footnote{https://physics.nist.gov/PhysRefData/ASD/lines\_form.html}. 
Due to the high wind densities, and the decrease in
ionization with radius, the inclusion of these additional
transition probabilities did not signifcantly change the
predicted spectrum.

Although the data were dereddened as described above,
it is very difficult to derive an accurate extinction
by fitting the optical-UV spectrum,
since a ``continuum''  is not observed -- most of the spectrum
is dominated by broad overlapping emission features. The overlap
of the emission features is exacerbated by the large terminal
velocity of the wind, $\vinf \sim  15\,000$\vunit.


\begin{figure*}
 \centering
 \includegraphics[width=0.45\textwidth]{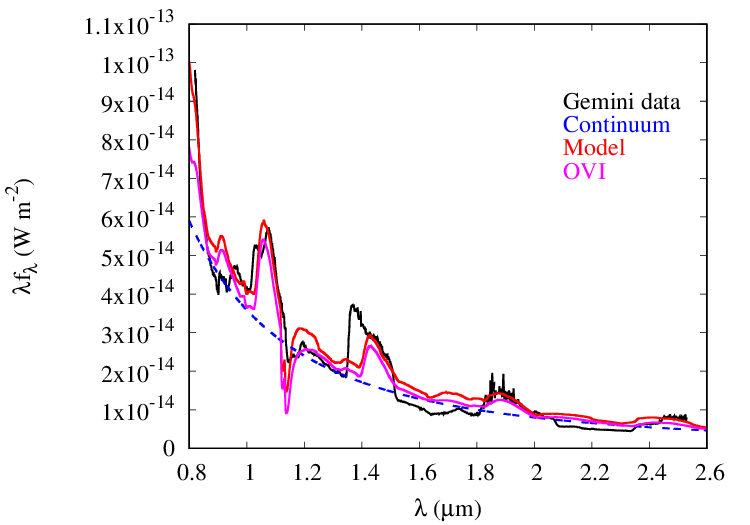}
  \includegraphics[width=0.45\textwidth]{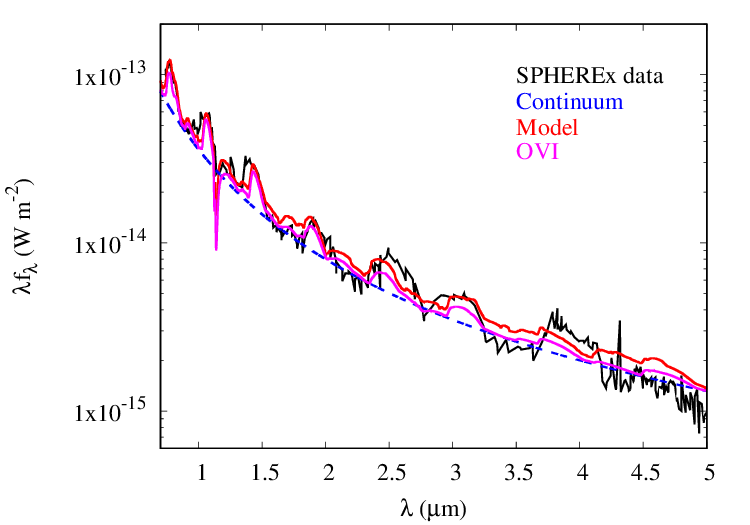}
 \includegraphics[width=0.45\textwidth]{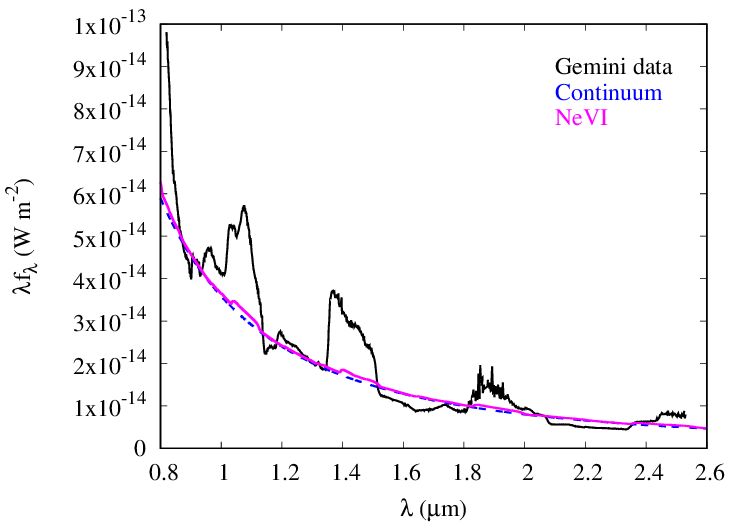}
 \includegraphics[width=0.45\textwidth]{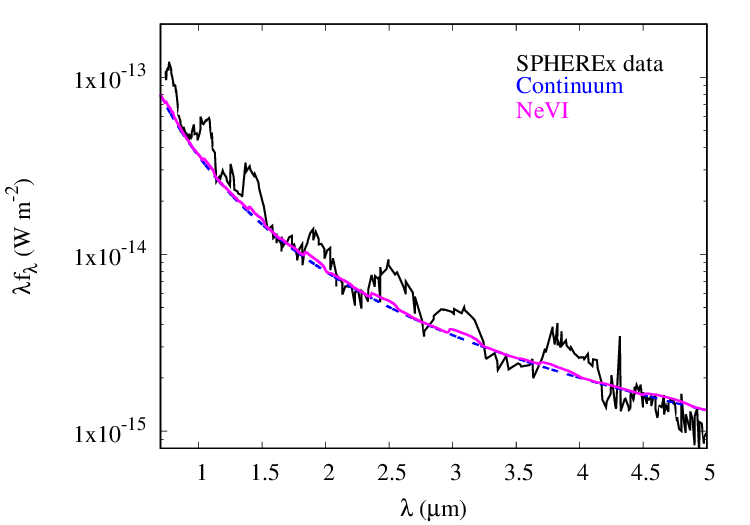}
 \caption{Comparison of CMFGEN model as described in text.
 Top: model based on \protect\cite{fesen23}, covering 
 the Gemini data (left) and the SPHEREx data (right).
 The continuum is shown as a blue dotted line, and the model
 spectrum is shown in red. The sum of \ion{O}{vi}
 and the continuum is also shown (magenta).
Bottom: as top but for model based on 
\protect\cite{fink14}. The continuum is again shown as a
blue dotted line; the sum of \ion{Ne}{vi} and the 
continuum is also shown in magenta.
\label{cmfgen_o}}
\end{figure*}

The model is compared with the Gemini data
(0.8\mic--2.5\mic), and the SPHEREx data (0.7\mic--5\mic)
in Fig.~\ref{cmfgen_o}. It is evident that the model provides
a fairly close fit to the higher resolution Gemini data,
and to the lower resolution, but greater wavelength range,
of the SPHEREx data.

The dominant features in the model near IR spectrum of Pa~30
belong to \ion{O}{vi}, and in particular transitions of the 
form $n \rightarrow n-1$.
For \ion{O}{vi}, $s, p, d$, and $f$
states have distinct energies, while states with $l=5$ 
(i.e. $g$ and above) and higher have virtually identical
energies. In our model atoms these high-$l$ states are
generally grouped into a single level (and given the 
designation $z$). For even higher principal quantum numbers,
all $l$ states of a given multiplicity are combined and given
a $W$ designation. The wavelengths of the near IR transitions
are generally unknown --- for high-$l$ states the energy
levels were computed assuming a quantum defect of 
zero\footnote{NIST lists energies for some states, but they
are uncertain and tend to have variable quantum defects.}.

The P~Cygni profile with its absorption near 1.12\mic\
is due to \ion{O}{vi}
$3p$ $^2$P$^{\rm o}$ --  $3d$ $^2$D. The lower level of this
transition gives rise to the intense \ion{O}{vi} emission
near 3820\,\AA\ \citep[see Figure~2 of][]{lykou23}.
The emission lines associated with \ion{O}{vi}
$3p$ $^2$P$^{\rm o}$ -- $3d$ $^2$D are prominent in
Fig.~\ref{cmfgen_o}, but are much weaker than the 
optical emission lines since the  $3d$ $^2$D level prefers (by a 
factor in excess of $6 \times 10^4$)
to decay to $2p$ $^2$P$^{\rm o}$.

Many other hydrogenic-like transitions  (ie., $n \rightarrow n-1$)
from \ion{C}{iv}, \ion{O}{v},  \ion{Ne}{iv}, \ion{Ne}{v}, and
\ion{Ne}{vi}, are also present in the near-IR; in our model their
peak intensities are typically less than 15\%, and are not explicitly
shown in Fig.~\ref{cmfgen_o}. 

Our model atoms typically include levels up to $n=30$, and
no correction for higher lying levels is made. This means
that the strengths of our IR transitions are underestimated.
The correction should generally be less than 20\% for the
transitions of interest.
However the correction for \ion{O}{v} is likely to be much
larger -- potentially by an order of magnitude. In our model,
emission from \ion{O}{v} originates at temperatures of 
around 20,000\,K. At this temperature dielectronic 
recombination is important for \ion{O}{v}. Dielectronic
recombination occurs through states with high principal
quantum number (in this case $2p$ $nl$ states that converge
on \ion{O}{vi} $2p~^2$P$^{\rm o}$). Assuming a hydrogenic
approximation, the radiative recombination rate coefficient
into O$^{4+}$  states with $n=10$ and higher  is approximately
$3 \times 10^{-12}$\,cm$^3$\,s$^{-1}$, while the dielectronic
recombination rate coefficient at 20000\,K is 
$\approx 3 \times 10^{-11}\,$cm$^{3}$\,s$^{-1}$
\citep[computed using the fitting formula given by][]{mazzotta98}. 
Thus the dielectronic recombination rate dominates. 

As the progenitor of Pa~30 is not well understood, we also
considered alternative models having different abundance sets.
In particular, we adopted the abundance sets of 
\cite{fink14} and \cite{lach22},
although we changed the C, O, and Ne abundances to retain
agreement with the UV/optical spectrum. In general rough
agreement could be obtained with the observations, but 
none had noticeably superior agreement than the ``standard''
model described above. In the high Fe abundace model from
\cite{fink14}, a higher luminosity is required
to produce the observed strength of \ion{O}{vi} 
$\lambda\lambda$ 3811, 3834\AA. Due to the higher $L$ and the
higher Fe abundance, our model has the distinct advantage
that radiation pressure acting on Fe could potentially 
drive the observed flow. However the model tended to produce
too much continuum flux in the IR, and lower emission equivalent
widths of the near IR \ion{O}{vi}\ lines. A much larger grid
of models would be needed to produce a more realistic 
determination of the model set (with a full parameter 
exploration) that provides the best match to Pa~30.

\begin{figure}
 \includegraphics[width=0.45\textwidth]{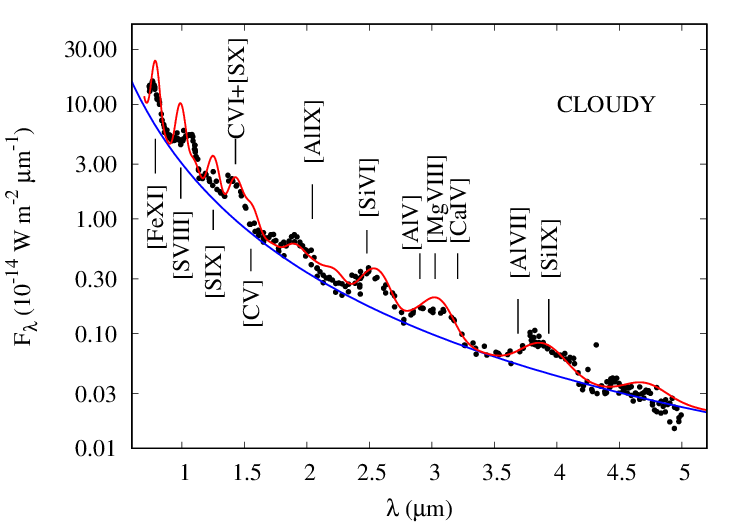}
 \includegraphics[width=0.45\textwidth]{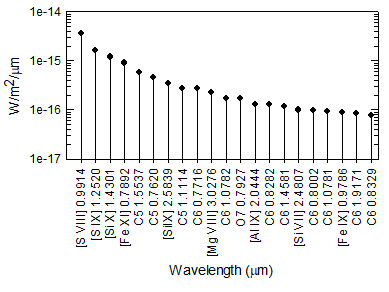}
 \caption{Top: CLOUDY model fit to SPHEREx data. Black dots:
SPHEREx data, dereddened by $E(B-V) = 1.05$. Blue curve:
model continuum, red curve: continuum plus emission lines
superposed. Bottom: line fluxes of $\sim20$ of the strongest
lines, in descending order, that contribute to  the CLOUDY
model of  the top panel. \label{fit}}
\end{figure}

\subsection{{CLOUDY modelling of Pa30}}

\begin{table}
\caption{CLOUDY coronal equilibrium model at temperature
$10^6$~K.\label{cloudy}}
 \begin{tabular}{ll}
 Parameter & Status/value \\ \hline
Photoionization  & Off \\
Stellar radius$^*$ & 0.155\Rsun \\
Geometry   & Thin spherical shell \\
Inner radius $R_{\rm in}$ &  $10^{13}$~cm \\
Shell thickness $R_{\rm out} - R_{\rm in}$ & 
$10^{10}$~cm \\
Wind terminal velocity $v_\infty$ & 15,000\vunit \\
Electron density  & $n_e \propto r^{-3}$ \\
Electron density & $3.86\times10^{10}$~cm$^{-3}$ at $R_{\rm in}$ \\
Distance & 2.1 kpc \\ \hline
\multicolumn{2}{l}{$^*$From \cite{gvaramadze19}.} \\
 \end{tabular}
\end{table}

\begin{table}
\centering
\caption{Elemental abundances adopted in the CLOUDY model.}
\label{tab:massfractions}
\begin{tabular}{llll}
\hline
Element & Mass Fraction & Element & Mass Fraction \\ \hline
H  & $1.0\times10^{-6}$ & Si & $1.44\times10^{-2}$\\
He & $1.48\times10^{-5}$ & P  & $1.433\times10^{-4}$\\
C  & $4.079\times10^{-1}$ & S  & $3.848\times10^{-3}$ \\
N  & $1.56\times10^{-4}$ & Cl & $1.776\times10^{-5}$\\
O  & $4.675\times10^{-1}$ & Ar & $6.199\times10^{-4}$\\
Ne & $5.20\times10^{-2}$ & Ca & $3.357\times10^{-4}$\\
Na & $1.645\times10^{-3}$ & Fe & $2.19\times10^{-3}$\\
Mg & $2.49\times10^{-2}$ & Ni & $2.674\times10^{-3}$\\
Al & $1.905\times10^{-3}$ \\ \hline
\end{tabular}
\end{table}

CMFGEN, based on the \cite{fesen23} parameters,
satisfactorily reproduces the Gemini and SPHEREx spectra,
using primarily emission lines of O, C, and Ne. However
several of the SPHEREx spectral features coincide with 
coronal lines of S, Si, Mg, and Al (hereafter referred to
as intermediate mass elements; IMEs), often seen in the
spectra of novae \citep[see, e.g.,][]{woodward21,kumar22}. 
The X-ray spectrum of the central star of Pa30 
in \cite{oskinova20} contains
several lines of  Ne, Mg, Si, S, and Fe.
Their study showed that the central star and its nebula are
luminous X-ray sources, so coronal gas at $\sim10^6$~K is to
be expected. Further, the early-time spectra of SNe Iax
contain Fe-group and IMEs (including
Si, S, and Ca), showing that these elements have been
brought to the surface during the explosion 
\citep[see][and references therein]{jha17}. 
The presence of coronal emission from IMEs is therefore
expected, and we used CLOUDY \citep[C25.0;][]{gunasekera25}
to explore this possibility.
We restrict the present study to the collisionally
ionised coronal plasma case, which successfully accounts
for the observed  high-ionization emission lines.

We adopt a collisionally ionised coronal plasma with
a temperature of $10^{6}$~K; the CLOUDY input parameters
are listed in Table~\ref{cloudy}.
The assumed elemental mass fractions are listed in 
Table~\ref{tab:massfractions}, in which
the abundances are those expected in the bound remnants
of SNe Iax \citep{fink14,lach22}; the abundances assumed
are for a bound remnant with mass 1.03\Msun\
\citep[see model N5def in Tables~3 and B3 of][]{fink14}, 
but with the Fe abundance reduced by a factor of 10.
The Pa30 ejecta are almost completely devoid
of H and He. The hydrogen number density at the 
shell was taken to be
$10^{5}\,\mathrm{cm^{-3}}$, resulting in an 
electron density of $\sim4\times10^{10}\,\mathrm{cm^{-3}}$.
The severe depletion of H and He means that 
free electrons are predominantly supplied by ionised
heavy elements rather than by H.

The velocity field is described by
\begin{equation}
v(r)=v_{\infty}\left(1-\frac{R_{\star}}{r}\right) ,
\end{equation}
where $v_{\infty}=15,000\ \mathrm{km\,s^{-1}}$ is the terminal
velocity.
For $r=10^{13}$ cm, the term $R_{\star}/r \ll 1$; therefore 
$v(r)\approx{v}_\infty$.

The CLOUDY model results are presented in Fig.~\ref{fit}
(top panel), which shows a synthetic spectrum
constructed using the fluxes of the 250 strongest
lines in the range 0.5--5\mic\ as listed by CLOUDY of the
elements listed in Table~\ref{tab:massfractions}.
Each line was assumed to be a gaussian with a full
width at half maximum (FWHM) of 
$\sim24,000$\vunit\, corresponding to the measured FWHM of
$\sim300$\AA\ for the \ion{O}{vi} 3,811\AA, 3,834\AA\ lines
\citep{gvaramadze19}. The spectrum was constructed over a
grid of wavelength points between 0.7--5.0\mic\ by a 
convolution of all the gaussians after scaling for their
peak intensities using their observed line fluxes. 
The emission line spectrum was superimposed on the 
continuum supplied by the CMFGEN model (see above).
It is evident that several coronal 
lines of Si, S,  Mg, Al, and Fe 
match the features in the SPHEREx spectrum, suggesting
that they could be potential contributors to the observed
spectrum. This would be consistent with the work of
\cite{oskinova20}, who estimate  even higher mass fractions
for Mg, Si and S for the central star 
($X(\mbox{Mg}) = 0.04\pm 0.02$; $X(\mbox{Si})
= 0.06\pm 0.04$, $X(\mbox{S}) = 0.04$)
rather than those given in Table~\ref{tab:massfractions}.

\section{{Summary}}

We have reported and analysed the first near-infrared
spectra of the central star of Pa30. Modeling with
CMFGEN indicates that most of the spectral features
can be explained as arising from \ion{O}{vi}.
Furthermore, an exploratory analysis with CLOUDY suggest
there is a region further out in the wind than the
region of \ion{O}{vi} line emission which is hot and
shock-heated, giving rise to coronal line emission from IMEs.
The observed spectrum of Pa30 is likely a superposition of
both these spectral components.

Many lines of S, Si, Mg, Al etc. are present in the
mid-IR region covered by the James Webb Space Telescope
(JWST) \citep[e.g., Table~2 of][]{evans12}.
Likely planned JWST spectra may establish the
relative contributions of the two components.
The case of SN~2024vjm underlines
how powerful ground-based and JWST observations can be
for studying radioactive products, slow ejecta, dust
formation, and possible signatures of a surviving remnant
\citep{kwok25,zimmerman26}. A JWST spectrum would be 
especially useful to separate the broad C/O/Ne wind
component from any coronal contribution by IMEs.

\section*{Acknowledgements}

{We thank the referee for helpful comments.}
This publication makes use of data products from the
Spectro-Photometer for the History of the Universe, Epoch of
Reionization and Ices Explorer (SPHEREx), which is a 
joint project of the Jet Propulsion Laboratory and the
California Institute of Technology, and is funded by the
National Aeronautics and Space Administration. 
We thank Sean Bryan and Pao-Yu Wang for their
invaluable advice and help with the SPHEREx data.

The Gemini observations were made possible by award of
time for programme
{FT-2026A-140}. The international Gemini Observatory
is a program of NSF's NOIRLab, which is managed by the
Association of Universities for Research in Astronomy
(AURA) under a cooperative agreement with the National
Science Foundation, on behalf of the Gemini Observatory
partnership: the National Science Foundation 
(United States), National Research Council (Canada), 
Agencia Nacional de Investigaci\'{o}n y Desarrollo (Chile), 
Ministerio de Ciencia, Tecnolog\'{i}a e Innovaci\'{o}n (Argentina), 
Minist\'{e}rio da Ci\^{e}ncia, Tecnologia, 
Inova\c{c}\~{o}es e Comunica\c{c}\~{o}es (Brazil), 
and Korea Astronomy and Space Science Institute (Republic of Korea).

This paper is also based in part 
on observations made with the Italian Telescopio
Nazionale Galileo (TNG) operated on the island of La 
Palma by the Fundaci\'{o}n Galileo Galilei of the INAF
(Istituto Nazionale di Astrofisica) at the Spanish
Observatorio del Roque de los Muchachos of the Instituto
de Astrofisica de Canarias.
We acknowledge the scientific and technical staff of the 3.6~m
Devasthal Optical Telescope (DOT), a national facility operated and
managed by the Aryabhatta Research Institute of Observational Sciences
(ARIES), an autonomous institute under the Department of Science and
Technology, Government of India, for their support in facilitating
service-mode observations and providing the observational data.

DPKB and VJ acknowledge support from PRL, Ahmedabad
and the Department of Space, Government of India.
SS acknowledges partial support from a NASA
Emerging Worlds grant to ASU (80NSSC22K0361) as well as
support from his ASU Regents' Professorship and 
JWST/NASA.

The research by DJH was supported in part by the University
of Pittsburgh Center for Research Computing and Data, 
RRID:SCR\_022735, through the resources provided. 
Specifically, this work used the H2P cluster, which is
supported by NSF award number OAC-2117681.

CEW acknowledges partial support from NASA JWST-06866.022-A.
TRG's research is supported by the international Gemini
Observatory, a program of NSF NOIRLab, which is managed by the
Association of Universities for Research in Astronomy (AURA)
under a cooperative agreement with the U.S. National Science
Foundation, on behalf of the Gemini partnership of Argentina,
Brazil, Canada, Chile, the Republic of Korea, and the United
States of America. 
GS acknowledges the WOS-A grant from the Department
of Science and Technology, India (SR/WOS-A/PM-2/2021).

\section*{Data availability}

The SPHEREx data used in this manuscript can be found
at 10.26131/IRSA629 SPHEREx Quick Release Spectral Images, 
https://doi.org/10.26131/IRSA629. 

The raw Gemini data are available from the Gemini Observatory
Archive, https:// archive.gemini.edu/.
The TNG and DOT data will be made available on request
to the corresponding author.


\bsp	
\label{lastpage}
\end{document}